\documentclass[twocolumn]{aastex631}
\usepackage{amsmath}
\usepackage[normalem]{ulem}
\usepackage{float}

\usepackage{times}
\usepackage[varg]{txfonts}

\begin{document}

\title{Impact of higher harmonics of gravitational radiation on the population inference of binary black holes}

\author[0000-0001-8081-4888]{Mukesh Kumar Singh}
\affiliation{International Centre for Theoretical Sciences, Tata Institute of Fundamental Research, Bangalore 560089, India}

\author[0000-0001-5318-1253]{Shasvath J. Kapadia}
\affiliation{International Centre for Theoretical Sciences, Tata Institute of Fundamental Research, Bangalore 560089, India}
\affiliation{Inter University Centre for Astronomy and Astrophysics, Pune - 411007, India}

\author[0000-0002-4103-0666]{Aditya Vijaykumar}
\affiliation{Canadian Institute for Theoretical Astrophysics, University of Toronto, 60 St George St, Toronto, ON M5S 3H8, Canada}
\affiliation{International Centre for Theoretical Sciences, Tata Institute of Fundamental Research, Bangalore 560089, India}

\author[0000-0001-7519-2439]{Parameswaran Ajith}
\affiliation{International Centre for Theoretical Sciences, Tata Institute of Fundamental Research, Bangalore 560089, India}
\affiliation{Canadian Institute for Advanced Research, CIFAR Azrieli Global Scholar, MaRS Centre, West Tower, 661 University Ave, Toronto, ON M5G 1M1, Canada}

\begin{abstract}
Templates modelling just the dominant mode of gravitational radiation are generally sufficient for the unbiased parameter inference of near-equal-mass compact binary mergers. However, neglecting the subdominant modes can bias the inference if the binary is significantly asymmetric, very massive, or has misaligned spins.  In this work, we explore if neglecting these subdominant modes in the parameter estimation of non-spinning binary black hole mergers can bias the inference of their population-level properties such as mass and merger redshift distributions. Assuming the design sensitivity of the advanced LIGO-Virgo detector network, we find that neglecting subdominant modes will not cause a significant bias in the population inference, although including them will provide more precise estimates. This is primarily because asymmetric binaries are expected to be rarer in our detected sample, due to their intrinsic rareness and the observational selection effects. The increased precision in the measurement of the maximum black hole mass can help in better constraining the upper mass gap in the mass spectrum. 
\end{abstract}

\section{Introduction}\label{sec:introduction}
The detection of gravitational waves (GWs) from the mergers of compact binaries has opened a new window of astronomy. By the end of the third observing (O3) run of LIGO-Virgo-KAGRA (LVK) detector network \citep{advligo, advvirgo, KAGRA}, we have observed $\sim 90$ GW signals from merging compact binaries \citep{gwtc-1, gwtc-2, gwtc-3, ias-1, ias-4, ias-3, ias-2, Nitz_OGC1, Nitz_OGC2, Nitz_OGC3, Nitz_OGC4}. The observation of the electromagnetic (EM) counterparts of a GW signal from a binary neutron star (BNS) merger marked the beginning of multimessenger astronomy with GWs \citep{GW170817-DETECTION, GW170817-MMA}. The catalog of detections also includes an additional BNS merger and two neutron star-black hole (NSBH) mergers with no observed EM counterpart \citep{gw190425, nsbh_det_LIGO}. However, the vast majority of the observed signals correspond to binary black holes (BBHs).
 
From these observations of compact binary mergers, we have also inferred their intrinsic (masses and spins of the compact objects) and extrinsic (distance, orientation, sky-location, etc., of the binary) parameters \citep{GW150914_PE}. These observations suggest the existence of a new population of heavier and mildly spinning black holes (BHs) \citep{GW190521, population_implications_gw190521}. This is in contrast to lighter masses and relatively high spin of BHs observed in X-ray binaries \citep{x_ray_BH_mass, x_ray_BH_spin}~\footnote{\cite{Fishbach_GW_X_ray_bhs} show that there is no evidence of the BH mass distributions inferred using GWs and X-rays being different if one accounts for selection effects correctly in both scenarios. However, the spin distributions of BHs from GW and X-ray observations are in tension at $>99.9\%$ level.}. Determining the formation channel(s) of these merging BBHs is thus an area of active study \citep{formation_channels_bbh_mike2, formation_channels_gerosa, formation_channels_bbh_mike, formation_channels_stevenson}. A key step to constraining the formation channels is the inference of population properties of BBHs, which can be achieved by estimating the parameters that govern the shape of the distributions of the source parameters \citep{mandel_farr_gair, talbot_thrane}. 

Recently, the LVK collaboration has carried out population analyses using 76 compact binary mergers \citep{GW_pop_gwtc3} in the third GW transients catalog (GWTC-3). They inferred the underlying mass, spin, and redshift distributions of these binaries along with constraining the merger rates \citep{GW_pop_gwtc1, GW_pop_gwtc2, GW_pop_gwtc3}. Among various features, they found a peak in the BH mass spectrum near $\sim 34 M_{\odot}$ and a dearth of BHs beyond $\sim 60M_{\odot}$ potentially suggesting the presence of an upper mass gap associated with pulsational pair-instability supernovae (PPISNe) \citep{PISN_PPISN, PPISN} and pair-instability supernovae (PISNe) \citep{PISN_1, PISN_2, PISN_PPISN} respectively. The lack of BHs with masses below $\sim 6 M_{\odot}$ also points to the lower mass gap \citep{Farah_mass_gap}. The presence of these mass gaps in the mass spectrum is subject to the uncertainties caused by the limited number of detections. This analysis also observed a mild preference for positive aligned spins and a merger rate increasing with redshift. 

The inference of the true astrophysical population features depends not only on the flexibility of the population model chosen in the analysis but also on the accuracy of the estimates of the source properties. Any missing physics in the waveform models, such as the effect of subdominant modes, spin precession, orbital eccentricity, etc, can bias the inference of individual source properties that in turn will affect the inferred shape of the population distributions.

Neglecting subdominant modes of gravitational radiation can bias the parameter inference of compact binary mergers with significant asymmetries \citep{Varma2014, varma_ajith}. This can also lead to a loss of detections of BBH mergers when the higher modes are not included in GW searches \citep{Juan_HM_search, Capano_HM_search, Harry_HM_search, Divya_HMs}. Including higher modes in addition to the dominant mode in GW analyses provides unbiased and more precise measurements of the source parameters \citep{Chris_Anand_hm, Arun_HMs, imbh_hm_graff}, and can improve GW early-warning~\citep{Kapadia:2020kss, Singh:2020lwx, Singh:2022tlh}. An important reason for the improved precision in some of the parameters is that the higher modes also reduce correlations among them, such as luminosity distance and orbital inclination, initial phase and polarization angle, spin and mass ratio, etc. In particular, the better measurement of luminosity distance and inclination angle through higher modes would be useful in constraining cosmology \citep{Ish_h0_nsbh, cosmo_hms_LISA} and any off-axis beamed EM emission (e.g. short gamma-ray bursts) \citep{Arun_off_axis_grbs}. 

In this work, we study the effect of neglecting higher modes on the inference of population properties of non-spinning BBH mergers. Since the binary systems with approximate symmetry (near-equal component masses, aligned spins, etc.) are louder than the ones with significant asymmetry, the detected population is dominated by the former as a consequence of the Malmquist selection bias \citep{malmquist_bias1, malmquist_bias2}. Due to the beaming of gravitational radiation along the direction of the orbital angular momentum, there is also a selection bias towards face-on binaries, for which the effect of higher modes is minimal. This means including/neglecting higher modes in the analysis will not play a major role in the parameter estimation for most of the detected binaries that are nearly symmetric. However, even if neglecting higher modes does not cause a significant bias (the shift in the peak of the posterior from the true value is smaller than statistical uncertainties) in individual binaries, it might show up in the population-level parameters when accumulated over many sources. This is what we explore in this paper. 

As a simple model, we choose power-law models for masses and a source-frame merger rate that is uniform in comoving volume for simulating a non-spinning BBH merger population. We find that the detected population is dominated by near-symmetric and face-on BBH mergers. This does not lead to any significant bias in inferring the population properties when subdominant modes are neglected in the parameter inference of individual binaries. However, hyper-parameters describing the maximum mass of the BH and the merger rate evolution are measured more precisely when higher modes are included in the analysis. Also, when we choose a subpopulation of near-edge-on binaries, these hyper-parameters have more prominent biases, still smaller than the $90\%$ credible interval (CI). This indicates that if the fraction of events with significant higher modes contribution is large in the true population, it could bias the inferred estimates.

We have organized the rest of the paper as follows: In section \ref{sec:hm_multipole} we discuss the theoretical background of subdominant modes, followed by a brief introduction to population inference and selection effects. The results are presented in section \ref{sec:results} including parameter estimation of individual BBHs as well as population inference. Section \ref{sec:conclusion} summarizes the results and discusses the future directions.

\section{Method}\label{sec:hm_multipole}
\subsection{Higher order modes of gravitational radiation}
The gravitational radiation emitted by a compact binary merger is composed of two tensor polarizations, $+$ and $\times$, as predicted by general relativity (GR). GW amplitude $h(t)$ as a function of time $t$ can be written as a complex combination of these two polarizations:
\begin{equation}
    h(t) = h_+(t) - i h_{\times}(t).
\end{equation}
Furthermore, this combination can be expanded in terms of spin$-2$ weighted spherical harmonics~\citep{NewmanPenrose}
\begin{equation}
    h(t; \boldsymbol{\lambda}, \iota, \varphi_0) = \frac{1}{d_L}\sum_{\ell = 2}^{\infty} \sum_{m = -\ell}^{\ell} h_{\ell m}(t, \boldsymbol{\lambda}) Y_{-2}^{\ell m}(\iota, \varphi_0), 
\end{equation}
where $h_{\ell m}$ are various multipoles which depend on time $t$ and intrinsic parameters $\boldsymbol{\lambda}$ of the source viz. component masses ($m_1, m_2$) and spins ($\textbf{S}_1, \textbf{S}_2$) in case of a quasi-circular BBH system. The angular dependence of the radiation is captured by the basis functions $Y_{-2}^{\ell m}(\iota, \varphi_0) $, where  $\iota$ is the orientation of the orbital angular momentum vector of a binary with respect to the line of sight of the observer, and $\varphi_0$ is the initial reference phase defined as rotational offset of the detector frame compared to the source frame. The leading term in the expansion is the quadrupole mode ($\ell = |m| = 2$). The subleading terms in the expansion, i.e. terms with $\ell \geq 2$ and $|m| \neq 2$ are known as ``higher'' or ``subdominant'' or ``non-quadrupole'' modes. Theoretical templates modelling only the dominant quadrupole mode are generally sufficient to detect and infer source parameters of binaries that are nearly symmetric. However, neglecting higher modes could reduce detection efficiency and bias the inferred source parameters when the binaries have large asymmetries, e.g. low mass ratio $q = m_2/m_1$, large misaligned spins, etc. The orientation of binary, inclination angle $\iota$, also determines whether higher modes can be observed.  Systems that are not face-on ($\iota > 0$) will have a larger relative contribution of subdominant modes than near-face-on ($\iota \sim 0$) ones. 

For parameter inference in this work, we use the multipolar waveform approximant \texttt{IMRPhenomXPHM} \citep{XPHM} that has several subdominant modes [$(\ell, |m|) = (2, 1), (3, 3), (3, 2), (4, 4)$] apart from the dominant mode [$(\ell, |m|) = (2, 2)$], implemented in the \texttt{LALSimulation} library of \texttt{LALSuite} \citep{lalsuite}. We also want to compare parameter estimates obtained from this model to one obtained from a dominant-mode only model; for this, we use the  \texttt{IMRPhenomXP} \citep{XPHM} approximant.~\footnote{All these models incorporate the effect of precession but we always set the spin vectors to zero in our analysis.}

\subsection{Parameter inference}\label{sec:pe}
The detector strain time series can be written as a sum of the detector noise time series, and a GW signal known to exist in the chosen stretch of data and whose parameters are to be inferred:
\begin{equation}
    d = h  + n,
\end{equation}
where $h$ is the true GW signal and $n$ is the noise in the detector. Under the assumption that the noise $n$ is a stationary Gaussian random process, we can write the likelihood of the data $d$ given the source parameters ($\boldsymbol{\theta}$) of the GW signal as
\begin{equation}
    \mathcal{L}(d|\boldsymbol{\theta}) \propto \exp \left[ - \frac{1}{2} (d- h (\boldsymbol{\theta})|d -h(\boldsymbol{\theta})) \right],
\end{equation}
where $(.|.)$ denotes the inner product defined as
\begin{equation}
    (a|b) = 2 \int_{f_0}^{\infty} \frac{\tilde{a}(f) \tilde{b}^*(f) + \tilde{a}^{*}(f) \tilde{b}(f)}{ S_n(f)} \mathrm{d}f.
\end{equation}
The $S_n(f)$ is the power spectral density (PSD) of the noise and $f_0$ is the lower cut-off frequency corresponding to the sensitivity of the detector. The $\tilde{a}(f)$ and $\tilde{b}(f)$ are the Fourier transforms, and $*$ denotes the complex conjugate. The matched-filter signal-to-noise ratio (SNR) can be written in terms of the above-defined inner product as,
\begin{equation}
    \rho = \frac{(d|h)}{\sqrt{(h|h)}}.
\end{equation}
Assuming the prior information on the distribution of the parameters, i.e. $\pi(\boldsymbol{\theta})$, one can use Bayes' theorem to write the posterior on the parameters $\boldsymbol{\theta}$ given the data $d$
\begin{equation}
    p(\boldsymbol{\theta}|d) = \frac{\mathcal{L}(d|\boldsymbol{\theta})\pi(\boldsymbol{\theta})}{\mathcal{Z}(d)},
    \label{eq:bayes_theorem}
\end{equation}
where $\mathcal{Z}(d)$ is the normalization to the posterior probability distribution, also known as the evidence or marginalized likelihood
\begin{equation}
    \mathcal{Z} = \int \mathrm{d} \boldsymbol{\theta} \mathcal{L}(d|\boldsymbol{\theta})\pi(\boldsymbol{\theta}).
\end{equation}
Evidence is not essential for inferring the parameters but it is crucial when comparing two hypotheses (e.g. two theoretical models for the GW signal), say A and B, the ratio of evidences also known as Bayes' factor ($\mathcal{B}$) of model A with respect to B is:
\begin{equation}
    \mathcal{B}^{\rm{A}}_{\rm{B}} = \frac{\mathcal{Z}_A}{\mathcal{Z}_B}.
\end{equation}
When the two hypotheses have different prior probabilities for being true, a more sensible way of comparing models is the odds ratio
\begin{equation}
    \mathcal{O}^{\rm{A}}_{\rm{B}} = \frac{\mathcal{Z}_A}{\mathcal{Z}_B} \frac{\pi_A}{\pi_B},
\end{equation}
where the factor $\pi_A/\pi_B$ known as priors odds accounts for our prior belief in models chosen in the analysis. In this work, we do not give any preference for a multipolar over dominant mode waveform model before we carry out the inference, so we choose this ratio to be unity. The odds ratio will be the same as Bayes' factor in that case.
Computing posteriors $p(\boldsymbol{\theta}|d)$ is not an easy feat, especially when the dimensionality of the problem is large. For example, in the case of a BBH merger, the parameter space is 15-dimensional and it will be an impossible task to compute the probabilities on a grid. So one has to rely on stochastic sampling methods, such as Markov Chain Monte Carlo, nested sampling, etc., to compute the posteriors in a reasonable amount of time.

\subsection{Population inference}

Once we have a large number of GW events coming from an underlying source population, we can start estimating the properties of that population. This \emph{population inference} can be done in a hierarchical way: We first estimate the properties of individual sources as described in Sec \ref{sec:pe}. Using these posteriors and evidences from individual events, we infer the distribution of the population properties. 

Suppose we have observed $N$ GW signals denoted by the data set $\mathcal{D} \equiv \{ d_1, d_2, ..., d_N \}$, then the posterior probability distribution for the population hyper-parameters $\Lambda$, marginalized over the merger rate $\mathcal{R}$ \citep{loredo_selec_effects, mandel_farr_gair, pop_inf_salvo}, is given by 
\begin{equation}
    p( \Lambda| \mathcal{D}) \propto \frac{1}{p^N_{\rm{det}}(\Lambda)}\frac{\mathcal{L}( \mathcal{D}|\Lambda) \pi( \Lambda)}{\mathcal{Z}(\mathcal{D})},
    \label{eq:hyper_posterior}
\end{equation}
where $\mathcal{L}( \mathcal{D}|\Lambda)$ and $\pi( \Lambda)$ are the population likelihood (also known as hyper-likelihood) and prior (also known as hyper-prior) on hyper-parameters. 
The term $\mathcal{Z}(\mathcal{D})$ is the hyper-evidence that plays an important role when comparing two population models and is just a proportionality constant. In the above equation, $p_{\rm{det}}(\Lambda)$ encodes detector selection effects (see Sec.~\ref{subsec:selection_effects} for details). Under the assumption that different observations are independent of each other, we can write the population likelihood as:
\begin{equation}
    \mathcal{L}( \mathcal{D}|\Lambda) = \prod_{i=1}^{N} \mathcal{L}(d_i| \Lambda) = \prod_{i=1}^{N} \int \mathrm{d}\boldsymbol{\theta_i} \, \mathcal{L}(d_i| \boldsymbol{\theta_i}) \, \pi(\boldsymbol{\theta_i}|\Lambda),
    \label{eq:combined_likelihood}
\end{equation}
where $d_i$ denotes the data measurement for $i^{th}$ event which has its own set of parameters $\boldsymbol{\theta_i}$. Here $\pi(\boldsymbol{\theta_i}|\Lambda)$ denotes the population model prior for the distribution of $\boldsymbol{\theta}$ conditioned upon the hyper-parameters $\Lambda$. This assumes that all detected events have been drawn from the same population model. Now, the hyper-posterior upon substitution from Eq. \ref{eq:combined_likelihood} is given by
\begin{equation}
    p( \Lambda| \mathcal{D}) \propto \frac{\pi( \Lambda)}{p^N_{\rm{det}}(\Lambda)} \prod_{i=1}^{N} \int \mathrm{d}\boldsymbol{\theta_i} \ p(\boldsymbol{\theta_i}| d_i) \frac{\pi(\boldsymbol{\theta_i}|\Lambda)}{\pi(\boldsymbol{\theta_i})},
    \label{eq:hyper_pos_selec_effects}
\end{equation}
where we have used Bayes' theorem, Eq. \eqref{eq:bayes_theorem}, to write the likelihood for individual BBH events in terms of the posterior, prior, and evidence. The multiplication of evidences $\prod_{i = 1}^{N} \mathcal{Z}(d_i)$ is absorbed in the proportionality as it is constant. It is clear from the above equation that computing posteriors on hyper-parameters requires many evaluations of an integral which amounts to calculating the expectation value of the ratio of population prior to PE prior for each event in the catalog. The integral can be evaluated numerically via Monte-Carlo integration as:
\begin{equation}
    \int \mathrm{d}\boldsymbol{\theta_i} \ p(\boldsymbol{\theta_i}| d_i) \frac{\pi(\boldsymbol{\theta_i}|\Lambda)}{\pi(\boldsymbol{\theta_i})} \approx \frac{1}{N^i_{\rm{s}}} \sum_{j = 1}^{N^i_{\rm{s}}}  \frac{\pi(\boldsymbol{\theta_i^j}|\Lambda)}{\pi(\boldsymbol{\theta_i^j})}\Bigg\vert_{\boldsymbol{\theta_i^j}\sim p(\boldsymbol{\theta_i}| d_i)}
    \label{eq:monte_carlo_integration}
\end{equation}
where $\boldsymbol{\theta_i^j}$ denotes the $j^{th}$ sample drawn from the posterior of the $i^{th}$ event. The above approximation holds when there are a sufficient number of posterior samples. This number is dependent on the SNR of individual events and typically ranges $\mathcal{O}(10^3 - 10^5)$.

\subsection{Selection effects}
\label{subsec:selection_effects}
The detection of GW signals is limited by their loudness that depends on various parameters such as masses, spins, luminosity distance, orientation, the sky location (weakly), etc. This is known as the Malmquist bias \citep{malmquist_bias1, malmquist_bias2}. It is important to account for this bias as a correction factor in the population likelihood (notice the denominator of Eq. \eqref{eq:hyper_pos_selec_effects}) if one wants to understand the true astrophysics population properties of merging compact binaries. The detection probability for an astrophysical model with hyper-parameters $\Lambda$ can be computed by marginalizing the detection probability of binaries with various source parameters predicted by that model.
\begin{equation}
    p_{\rm{det}}(\Lambda) = \int \mathrm{d}\boldsymbol{\theta} p_{\rm{det}}(\boldsymbol{\theta}) \pi(\boldsymbol{\theta}| \Lambda),
\end{equation}
with 
\begin{equation}
    p_{\rm{det}}(\boldsymbol{\theta}) = \int_{d:f(d)\geq f_{\rm{th}}} \mathrm{d}d \ p(d | \boldsymbol{\theta}),
\end{equation}
where in the last equation, the integration is carried out only on those likelihoods for which data is such that it leads to the detection statistic $f(d)$ (e.g. matched-filter SNR or false alarm rate) greater than its threshold value $f_{\rm{th}}$. 
Computing $p_{\rm{det}}(\boldsymbol{\theta})$ numerically requires an extensive injection campaign of simulated GW signals in the detector noise followed by the calculation of the detection statistic. The finite number of simulations carried out introduces an uncertainty in the selection function estimated which must be mitigated by injecting enough synthetic GW signals \citep{selection_function_farr, selection_function_reed_farr, uncertainties_selec_func_colm, semianlytics_sensitivity_essick}.

\subsection{Population models}
\begin{table*}
    \centering
    \begin{tabular}{c | c | c}
        \hline 
        \hline
        Parameter & Description & Prior \\
        \hline 
        $\alpha$ & primary mass power-law spectral index & $U(-2, 4)$ \\
        $\beta_q$ & mass ratio power-law spectral index & $U(-3, 5)$ \\
        $m_{\rm{min}}$ & lower limit on the mass of the BH & $U(1, 15)$ \\
        $m_{\rm{max}}$ & upper limit on the mass of the BH & $U(60, 150)$ \\
        $\lambda_z$ & power-law spectral index of redshift distribution & $U(-6, 6)$ \\
        \hline
    \end{tabular}
    \caption{The population model hyper-parameters description and their corresponding priors chosen for hierarchical inference. Above, $U(a, b)$ indicates uniform distribution between $a$ and $b$. }
    \label{tab:hyper_param_prior_table}
\end{table*}
As a simple model, we choose a mass distribution model inspired by well-known initial mass functions of stars that are typically power laws \citep{salpeter}. In our simulations, we use a \textit{truncated} power-law model \citep{kovetz_model_B, maya_model_B}, known as ``Model-B'' in \cite{mass_model_A}~\footnote{{Note that the truncated power-law model is already in tension with the BH mass spectrum inferred from GW observations \citep{GW_pop_gwtc2, GW_pop_gwtc3}. For example, it can not explain the features in the mass spectrum at $\sim 35 M_\odot$.}}. The primary mass $m_1$ (the heavier BH in the BBH system) distribution with spectral index $\alpha$ is given by,
\begin{equation}
    p(m_1 | \alpha, m_{\mathrm{min}}, m_{\mathrm{max}}) \propto 
    \begin{cases}
    m_1^{-\alpha}, & m_{\mathrm{min}} < m_1 < m_{\mathrm{max}} \\
    0, & \rm{Otherwise},
    \end{cases}
    \label{eq:primary_mass_model}
\end{equation}
where $m_{\rm{min}}/m_{\rm{max}}$ is the lower/upper limit of the mass function. The distribution of the mass ratio $q \equiv m_2/m_1$ with power-law spectral index $\beta_q$ is,
\begin{equation}
    p(q | \beta_q, m_{\mathrm{min}}, m_1)  \propto
    \begin{cases}
    q^{\beta_q}, & m_{\mathrm{min}} < m_2 < m_1 \\
    0, & \rm{Otherwise}.
    \end{cases}
    \label{eq:mass_ratio_model}
\end{equation}
We assume the redshift distribution with power-law spectral index $\lambda_z$ that captures the evolution of the BBH merger rate with redshift\citep{merger_rate_model_pl}
\begin{equation}
    p(z) \propto  \frac{1}{1+z} \frac{dV_c}{dz} (1+z)^{\lambda_z}.
    \label{eq:redshift_model}
\end{equation}
With $\lambda_z = 0$, the above model corresponds to the distribution of BBHs uniform in 3-comoving volume and source frame time. The model hyper-parameters $\Lambda \equiv \{\alpha, \beta_q, m_{\rm{min}}, m_{\rm{max}}, \lambda_z \}$ are summarised in Table \ref{tab:hyper_param_prior_table} along with their priors chosen for hierarchical inference.

\section{Results}\label{sec:results}
We simulate an astrophysical population of 3900 non-spinning BBH mergers assuming that the primary mass $m_1$ is drawn from a log-uniform ($\alpha=1$) and the mass-ratio $q$ from a linear ($\beta_q=1$) distribution with minimum and maximum mass of a BH as $m_{\rm{min}}=5M_{\odot}, m_{\rm{max}}=100M_{\odot}$ respectively.
The BBH mergers are distributed uniformly in the source frame (uniform in comoving volume and source frame time) with redshift evolution hyper-parameter $\lambda_z = 0$. We model these GW signals using a multipolar waveform approximant, \texttt{IMRPhenomXPHM} \citep{XPHM}, a phenomenological frequency domain model implemented in \texttt{LALSimulation} \citep{lalsuite} and inject them into Gaussian noise, assuming the design sensitivity of the LIGO and Virgo detector network \citep{advligo, advvirgo}. 

\begin{figure}
    \centering
    \includegraphics[scale=0.33]{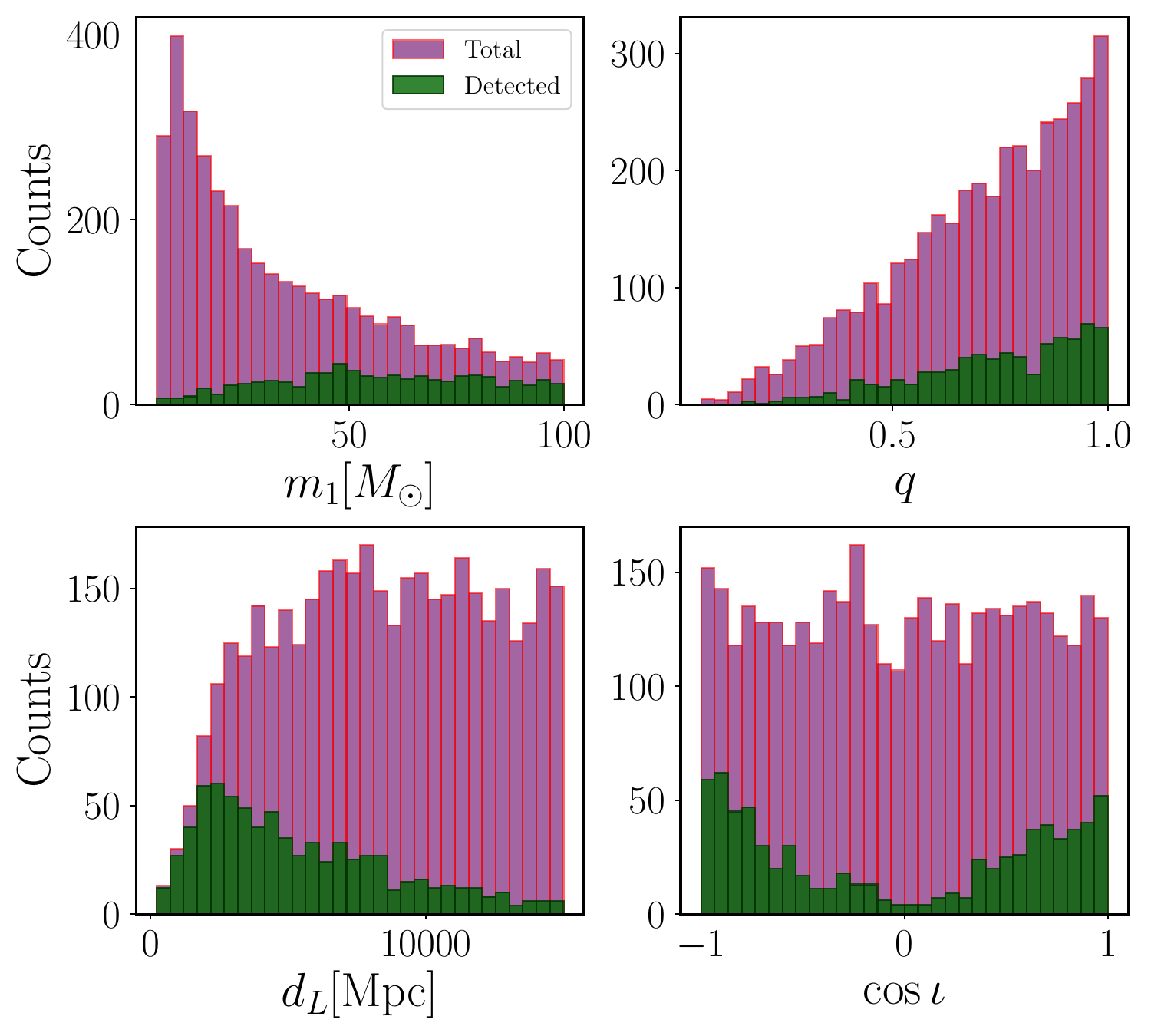}
    \caption{The distribution of the injected total population (in purple) of non-spinning BBH mergers along with the detected one (in green) that is computed assuming the network matched-filter SNR is greater than the threshold value of 8.}
    \label{fig:injected_detected_hist}
\end{figure}

We find $\sim 750$ injections were detected with a network matched-filter SNR threshold of 8. Fig. \ref{fig:injected_detected_hist} shows the distribution of injected and detected distribution of BBH mergers. We detect more massive and closer BBHs than lighter and farther away systems as expected by the selection effects. It is also noticeable that binaries with unequal masses and high inclination angles (for which the higher modes are expected to make a significant contribution) are rarer in the detected population. This is because asymmetric (small $q$) compact binary mergers are less efficient at emitting GWs than near-symmetric ones
\footnote{The optimal SNR of a binary is proportional to the square root of the symmetric mass ratio $\eta := q/(1+q)^2$~[see, e.g., Eq.~(B11) of \cite{Ajith:2007kx}]. Since the detection volume for a binary is proportional to the cube of the SNR, the fraction of detected events should roughly scale as ${\eta}^{3/2}$, which is consistent with what we see in Fig.~\ref{fig:injected_detected_hist}.}
Also, the radiation is primarily beamed along the direction of the orbital angular momentum. These effects will result in smaller SNRs, on average, for asymmetric and inclined binaries.

\subsection{Impact of higher modes on parameter inference}
\begin{figure*}
    \centering
    \includegraphics[scale=0.45]{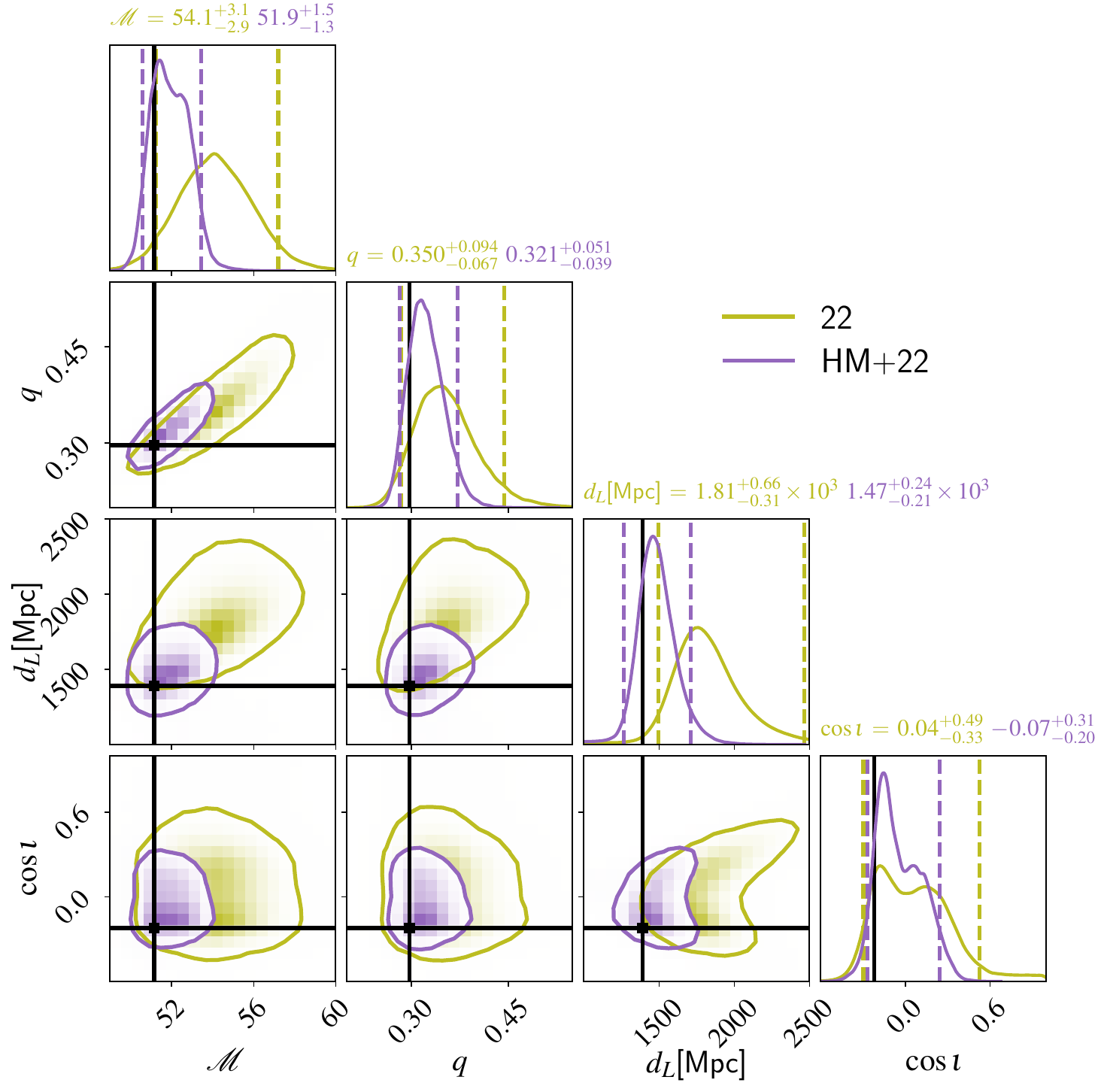}
    \caption{The two-dimensional along with one-dimensional posterior distributions on a few source parameters ($\mathcal{M}, q, d_L, \iota$) of a moderately asymmetric ($q\sim 0.3, \iota \sim 103$ degrees) BBH merger. Posteriors employing the higher modes are more accurate (peaking closer to the injected value, shown by the vertical black line) and precise (smaller spread in the posteriors; $90\%$ credible regions in the one-dimensional posteriors are shown by vertical dashed lines).  On the other hand, when the higher modes are neglected in parameter inference, it leads to not only the broadening of the posteriors but also the biased inference of the luminosity distance. The inclusion of higher modes has also reduced the degeneracy in inferring the source parameters, especially in $\iota-d_L$ and $\mathcal{M}-q$.}
    \label{fig:hm_vs_22_pe_example}
\end{figure*}

We performed parameter inference on these detected injections using both a dominant mode only (\texttt{IMRPhenomXP}) and multipolar (\texttt{IMRPhenomXPHM}) waveform model \citep{XPHM}. We used Bayesian Inference Library, \texttt{Bilby} \citep{bilby} that employs the \texttt{dynesty} \citep{dynesty_paper, dynesty_sampler} sampler to stochastically sample the posteriors on various parameters (masses, sky-location, luminosity distance, orbital inclination, polarization, and arrival time and phase) of the binaries. Fig. \ref{fig:hm_vs_22_pe_example} shows the two- and one-dimensional probability distributions of some of the important parameters (chirp mass $\mathcal{M}:= (m_1 m_2)^{3/5} / (m_1 + m_2)^{1/5}$, mass ratio $q$, inclination angle $\iota$, and luminosity distance $d_L$) after marginalizing over the rest. 
These marginalized distributions, plotted using the \texttt{corner} package~\citep{corner}~\footnote{The 2D probability contours that we compute will have some dependence on the bin size of the histogram: If the bin width is too large, it will not be able to resolve small differences between the posteriors from the ``22'' mode and ``HM+22'' parameter estimation. If the bin width is too narrow, this will create statistical fluctuations due to the limited number of samples per bin. We have chosen some intermediate bin width, with which the statistical fluctuations due to the limited number of posterior samples are insignificant. At the same time, the bin width is small enough so that we can see the difference between the two posteriors.}, correspond to a moderately asymmetric ($q\sim 0.3, \iota = 103$ degrees) BBH system recovered using \texttt{IMRPhenomXP} and \texttt{IMRPhenomXPHM}.
The parameter estimation is more accurate and precise when higher modes are included in the analysis. This shows the importance of higher modes in parameter estimation, consistent with the findings of many previous studies \citep{Chris_Anand_hm, Arun_HMs,  imbh_hm_graff, Varma2014, varma_ajith}. 

\begin{figure}
    \centering
    \includegraphics[scale=0.37]{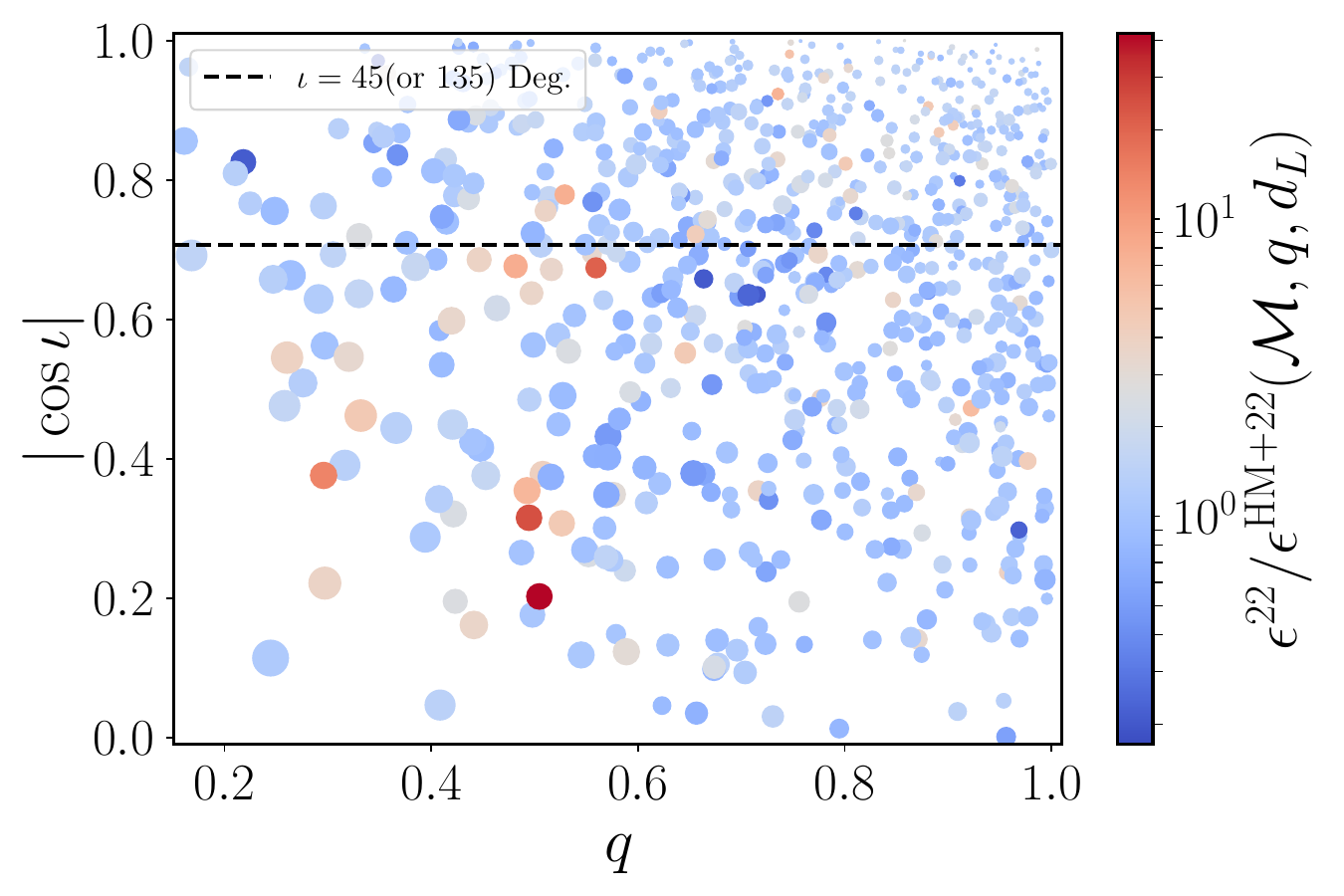}
    \caption{The ratio of the effective bias ($\epsilon^{\rm{22}} / \epsilon^{\rm{HM+22}}$) estimated for chirp mass, mass ratio, and luminosity distance recovery when using a multipolar waveform compared to a dominant mode waveform as a function of $q$ and $\iota$. The size of the marker corresponds to the SNR contribution due to the higher modes only.}
    \label{fig:quantile_mc_q}
\end{figure}

\begin{figure}
    \centering
    \includegraphics[scale=0.38]{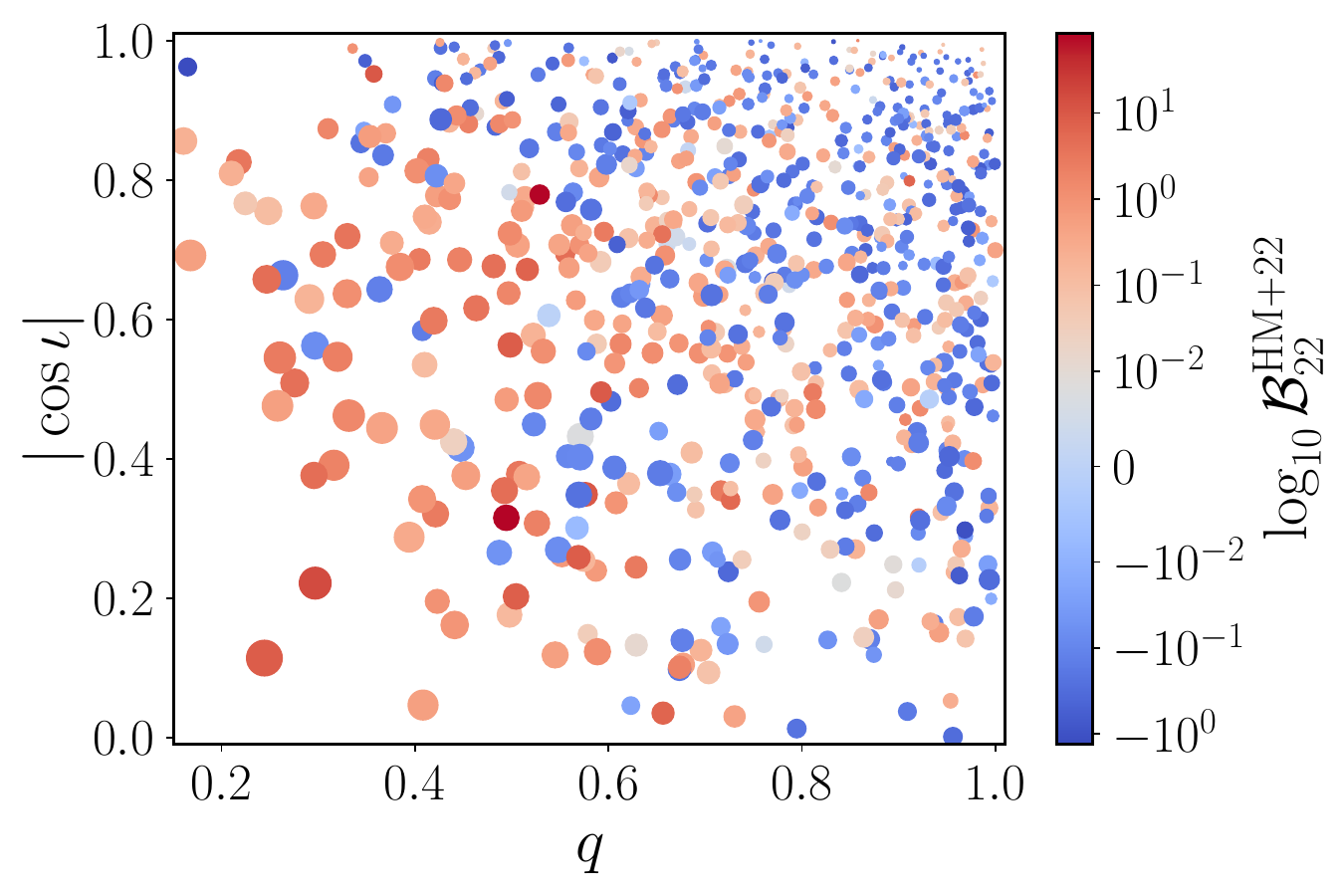}
    \caption{Same as Fig. \ref{fig:quantile_mc_q} but the color map represents the Bayes factors ($\mathcal{B}$) between a multipolar (HM+22) and only dominant mode (22) waveform models. In the asymmetric region of parameter space (low $q$ and non-face-on inclination), the multipolar waveform model is preferred over just the dominant mode.}
    \label{fig:log_BF_HM_22}
\end{figure}

To further quantify the effect of neglecting higher modes on the parameter inference of the BBH population, we compute a combined effective bias for the recovery of the chirp mass, mass ratio, and luminosity distance as 
\begin{equation}
\epsilon = \sqrt{\left(\frac{\hat{\mathcal{M}}}{\mathcal{M}^{\rm{tr}}} - 1 \right)^2 + \left( \frac{\hat{q}}{q^{\rm{tr}}} - 1 \right)^2 + \left(\frac{\hat{d}_L}{ d_L^{\rm{tr}}} - 1 \right)^2},
\end{equation}
where $\hat{}$ denotes the point estimates of the source parameters (median of the marginalized 1D posteriors in our case) and $\theta^{\rm{tr}}$ corresponds to the true values of the source parameters. We plot this effective bias $\epsilon$ as a function of $q$ and $|\cos\iota|$, the parameters that primarily determine the relative contribution of higher modes to the observed GW signal (see Fig. \ref{fig:quantile_mc_q}). The size of the marker represents the SNR contribution due to non-quadrupolar modes~\footnote{The SNR contribution due to HMs ($21, 33, 44$) is estimated by computing an orthogonal set of subdominant modes with respect to the dominant mode ($2$). If these orthgonal modes have the SNRs denoted by $\rho'_{\ell m}$, then the SNR contained in HMs is $\rho_{\rm{HM}} = ({\rho{'}_{\rm{21}}^2 + \rho{'}_{\rm{33}}^2 + \rho{'}_{\rm{44}}^2})^{1/2}$. The size of the marker in Fig. \ref{fig:quantile_mc_q} and \ref{fig:log_BF_HM_22} corresponds to the ratio $\rho_{\rm{HM}}/\rho_{\rm{22}}$ \citep{measuring_hms}.} which increases when moving from near equal masses and face-on systems to unequal masses and edge-on systems as expected. Note that the increase in the SNR is not monotonic as a function of the degree of asymmetry in the system. This is because the SNR contribution from higher modes for a given observing scenario is also determined by the source parameters such as total mass (spins also but we are assuming non-spinning binaries here) other than mass-ratio and inclination angle. The color-map density corresponds to the ratio of the effective biases ($\epsilon^{\rm{22}} / \epsilon^{\rm{HM+22}}$) estimated using a dominant mode only (22) waveform model to a multipolar (HM+22) one. The events with ratio $\epsilon^{\rm{22}} / \epsilon^{\rm{HM+22}} > 1$ denote that including higher modes improves the parameter estimation. We see an indication of a trend in the effective bias ratio which is mildly correlated with the SNR contained in the subdominant modes. The effective bias ratio is larger for more asymmetric events in general which means the higher modes play a significant role in inferring source parameters. We also plot the log-Bayes factor (ratio of the evidences) while comparing the parameter inference with a multipolar and dominant mode-only waveform model (see Fig. \ref{fig:log_BF_HM_22}). Again, the waveform model with subdominant modes is preferred over the dominant mode-only model when there is significant asymmetry in the binary system. 

\subsection{Inference of population properties}
\begin{figure*}
    \centering
    \includegraphics[scale=0.4]{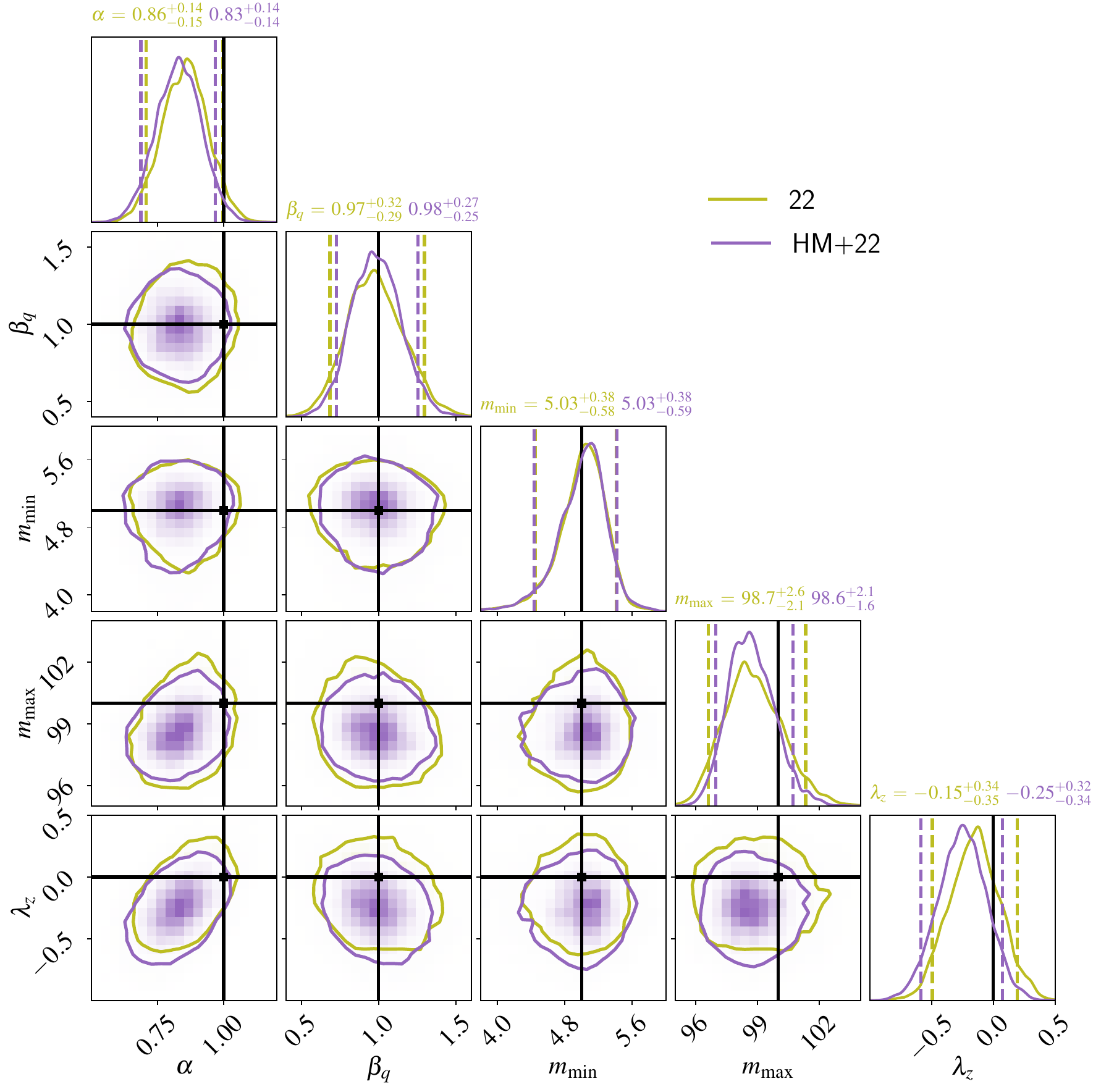}
    \caption{Two-dimensional posterior distributions along with the marginals of population hyper-parameters with (purple)/without (olive) using higher modes in the hierarchical inference of $750$ simulated GW events. Neglecting higher modes does not induce any significant bias in inferring the population properties but including them leads to more precise estimates of hyper-parameters, especially $m_{\rm{max}}$ and $\beta_q$. We also notice small biases in estimating $\alpha$ and $\lambda_z$ when higher modes are considered.}
    \label{fig:pop_hpe_posteriors_all}
\end{figure*}

\begin{figure*}
    \centering
    \includegraphics[scale=0.31]{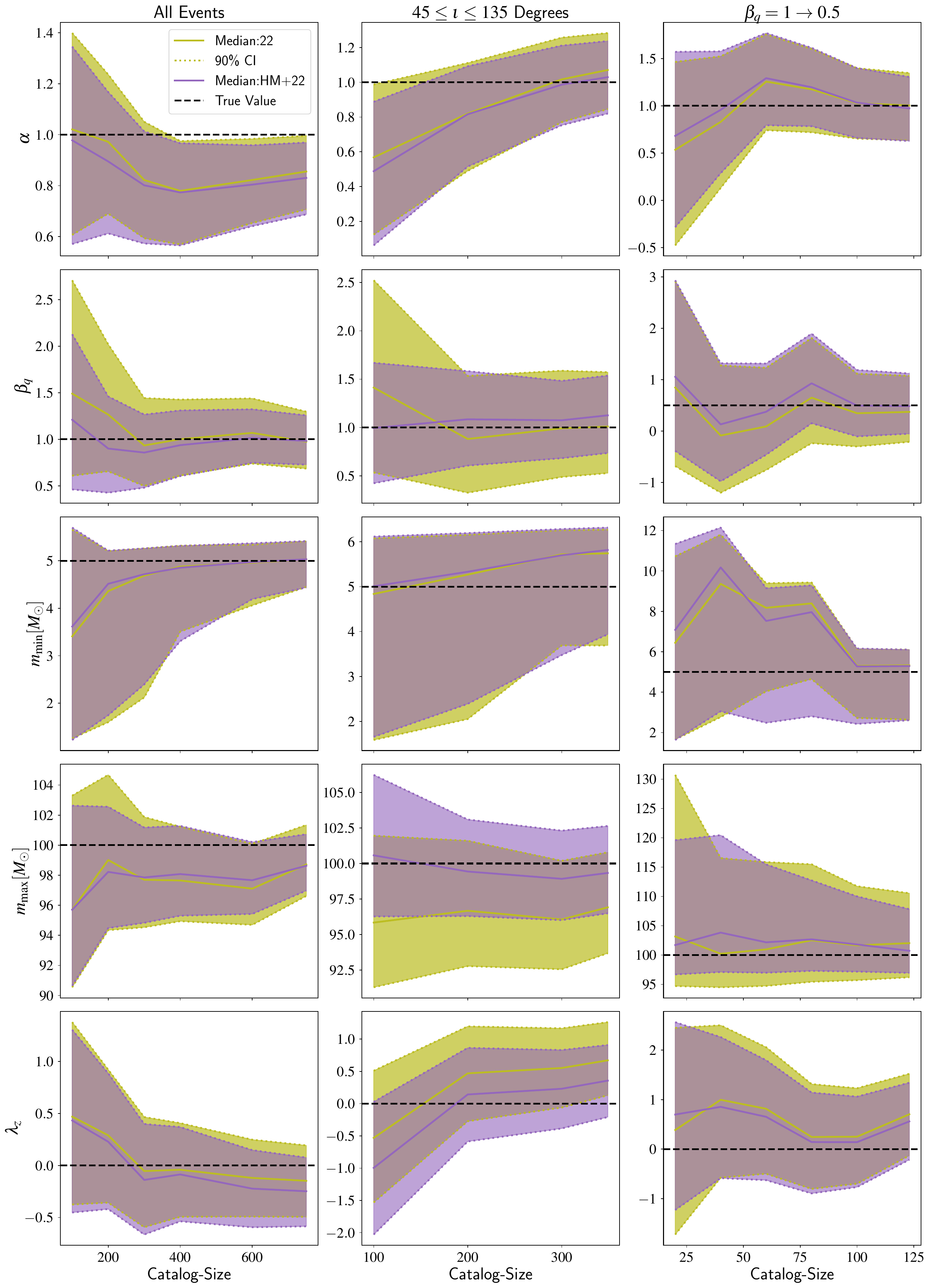}
    \caption{\textit{The left panel:} the $90\%$ width (between dotted lines) of the posteriors centered on the median values (solid lines) for various hyper-parameters as a function of catalog size (number of events in the catalog) when using all modes in the analysis (HM+22) along with only dominant mode (22) estimates. It is clear that there is some improvement in the statistical uncertainty in the hyper-posterior but we do not see any relative bias due to neglecting higher modes. \textit{The middle panel:} same thing as on the left but for injections with $45 \leq \iota \leq 135$ degrees. The increased fraction of events that have an observable contribution from higher modes leads to larger biases (still true value is within the $90\%$ confidence interval) for $m_{\rm{max}}$ and $\beta_q$ when higher modes are neglected. The hyper-parameter $\lambda_z$ is biased (at $90\%$ confidence interval) due to neglecting higher modes. On the other hand, the inclusion of higher modes leads to unbiased estimates of $\lambda_z$. \textit{The right panel:} Same as the left panel, except that the true value for $\beta_q$ is $0.5$ (as opposed to $\beta_q = 1$ in the left panel). This results in a larger fraction of asymmetric events (compared to the left panel), and still does not show any prominent biases. However, this might be limited by the smaller number of events ($\sim 120 $) in the catalog.}
    \label{fig:posterior_width_vs_catalog_size}
\end{figure*}

We consider posterior samples, from all the detected injections, obtained through nested sampling \citep{nested_sampling_skilling} using both multipolar (\texttt{IMRPhenomXPHM}) and dominant mode (\texttt{IMRPhenomXP}) waveform models. Specifically, this study focuses on the mass and redshift population properties of BBH mergers hence we only use posterior samples for masses ($m_1, q$) and redshift ($z$) in our analysis. We also compute a joint sampling prior (prior used during parameter estimation on individual events) $\pi(m_1, q, z)$ that is used to reweight the population prior $\pi(m_1, q, z|\Lambda)$ (the term inside the summation in Eq. \ref{eq:monte_carlo_integration}) in hierarchical inference. Further, the selection effects were estimated by injecting $50$ million GW signals into Gaussian noise and then computing the detection statistic, the network matched-filter SNR in our case, both when employing the multipolar and dominant mode waveform model. The found injections (GW signals with network matched-filter SNR $\geq 8$) were used to compute the selection effects \citep{selection_function_tiwari, selection_function_farr, selection_function_reed_farr}. In the case of real observations, this process becomes computationally expensive as it requires injecting signals in the detector noise and recording the detection statistics by running the search pipelines. 
In an ideal simulation, one should replicate the exact data analysis procedure used to detect GW signals in real data, which will have non-Gaussian tails. We leave this study for the future.

We infer the population properties of the mass and redshift of BBH mergers using a GPU-accelerated population inference code \texttt{GWPopulation} \citep{gwpopulation}. Fig. \ref{fig:pop_hpe_posteriors_all} shows the posteriors as well as 2D probability distributions on various hyper-parameters. We find that the population level parameters (especially $m_{\rm{max}}$, $\beta_q$, and $\lambda_z$) are more precisely measured when higher modes are included in the analysis. However, we do not observe any significant bias in inferring the hyper-parameters when higher modes are neglected. This can be understood from Fig. \ref{fig:quantile_mc_q}---the number of highly asymmetric (low $q$ and near-edge-on inclination) BBH systems is rather scarce and the population is dominated by near-symmetric BBH mergers. 

In Fig. \ref{fig:pop_hpe_posteriors_all}, note that while 1D posterior on hyper-parameter $\alpha$ presents a small bias for both models (dominant mode only and multipolar), it is recovered within the $90\%$ probability contours for 2D posterior distributions. Note that there is a $\sim10\%$ chance for the true value to be outside the 90\% credible region of the posterior. So, the fact that we are not recovering the true value of one parameter among six is not totally unexpected. However, we see a slightly larger bias in the multipolar posterior (as compared to the dominant mode). This could be an artifact of the Poisson fluctuations due to using only one catalog of events (one realization of the observing scenario). Due to the computational costs involved, we are unable to create a new catalog of simulated events. Hence we tested this hypothesis by doing the hyper-parameter estimation on different sub-catalogs (subsets) of events drawn from our existing catalog of 750 simulated events. We do see random fluctuations in the posteriors estimated from different sub-catalogs, consistent with our broad expectations. However, this issue needs to be investigated in detail in the future.

As the number of detections increases, we will get more precise estimates of population properties. We speculated that even though the effect of neglecting subdominant modes in the parameter inference may not cause any significant biases in the source properties of individual binary systems, it could bias the inference of population properties when accumulated over many events. We test this by estimating the errors in measuring the hyper-parameters as a function of the catalog size (the number of events detected). In the left panel of Fig. \ref{fig:posterior_width_vs_catalog_size}, we have plotted the $90\%$ width of the hyper-parameter posteriors centered on median values as a function of the catalog size. The inference of hyper-parameters (statistical uncertainty) becomes more precise when the number of detections grows. However, even with the accumulation of small biases over $\sim \mathcal{O}(700)$ events, we have not yet observed any significant bias in population-level parameters due to the neglect of subdominant modes in the analysis.

Note that the above conclusions are based on a single set of hyper-parameters used to generate the mock catalog of events. Ideally, one will have to repeat the population inference using different simulated values of hyper-parameters. Due to computational limitations, we leave this more exhaustive study for future work. However, we do partially investigate this by generating a population of simulated events corresponding to a new population model by selecting a subset of events from the original catalog using reweighting. We choose the mass-ratio hyper-parameter $\beta_q = 0.5$ with the rest of the hyper-parameter values unchanged for the new model. This model will produce a larger fraction of asymmetric events but also does not differ drastically from the old model. We find $\sim 120$ events detected for the new model out of $750$ events from the old model. In the right panel of Fig. \ref{fig:posterior_width_vs_catalog_size}, we again plot the $90\%$ CI of the hyper-posteriors along with median values as a function of catalog size. We do not find any strong indication of bias in the population inference when higher modes are neglected in the analysis.

\begin{figure*}[htbp!]
    \centering
    \includegraphics[scale=0.4]{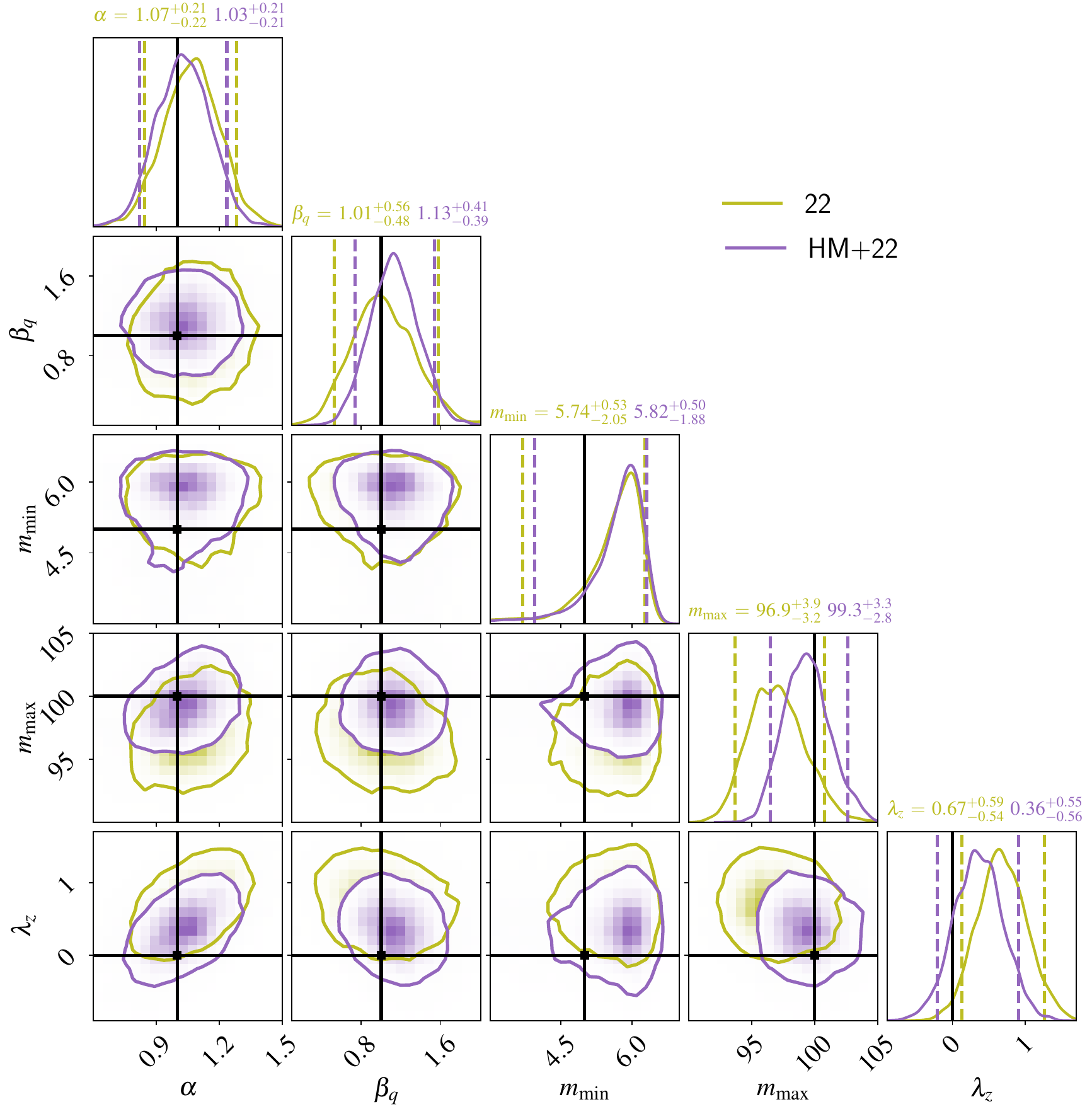}
    \caption{Same as in Fig. \ref{fig:pop_hpe_posteriors_all} but the injections with $45 \leq \iota \leq 135 $ degrees. Increasing the fraction of events with observable contribution from higher modes leads to larger bias in the estimates of hyper-parameters, especially, mass-ratio power-law spectral index $\beta_q$, maximum mass $m_{\rm{max}}$ and merger rate evolution hyper-parameter $\lambda_z$.}
    \label{fig:pop_hpe_posteriors_selec_iota_geq_30}
\end{figure*}

It will also be interesting to see how the bias in the population properties depends on the fraction of detected events that have a significant contribution from higher modes. Since more inclined systems will have a higher contribution from higher modes, we check if increasing the fraction of highly inclined binaries in the catalog changes the amount of bias. Specifically, we choose a subset of $\sim 350$ out of $750$ simulated GW events which have inclinations $45 \leq \iota \leq 135 $ degrees. Although a large fraction of the binaries are in this range, since highly inclined binaries are harder to detect than near-face-on configurations, the above inclination range includes only $\sim 47\%$ of total detected events.


Confining the sub-population within a specified inclination range changes the underlying astrophysical distribution of inclinations, and this should be accounted for while inferring population-level properties. For this purpose, we restrict the PE samples per event as well as injections for selection effects within the new inclination range to avoid any false biases. The hierarchical inference reveals notable biases in $\beta_q$ and $m_{\rm{max}}$ although the true value is within the $90\%$ CI when subdominant modes are neglected in the analysis (see Fig. \ref{fig:pop_hpe_posteriors_selec_iota_geq_30}). The hyper-parameter $\lambda_z$ is biased (at $90\%$ CI) when higher modes are neglected. However, including higher modes recovers the true value of $\lambda_z$. We check the stability of this bias by creating multiple realizations by drawing events randomly from the subpopulation with replacement and find that population inference recovers $\lambda_z$ (at $90\%$ CI) when including the effect of higher modes. The dominant mode analysis still leads to biased estimates. The hyper-parameters $m_{\rm{max}}$ and $\beta_q$ exhibit biases towards lower values when neglecting subdominant modes. 

The dominant mode templates corresponding to lower masses and more asymmetric systems will have a longer duration in an attempt to mimic a more extended signal with higher modes. Similarly, the bias in $\lambda_z$ towards larger values shows the loss of SNR due to neglecting subdominant modes in the analysis. It is noteworthy that the absence of higher modes in the parameter inference does not induce any bias in inferring the lower edge of BH mass spectrum $m_{\rm{min}}$, which can be explained by the fact that lighter systems are not expected to be asymmetric for higher modes to play any significant role. If the true population (in nature) has a significant fraction of detectable asymmetric events, the population properties will incur significant biases upon neglecting higher modes. Although we have not explored this, NSBH and massive binaries will be prone to this kind of systematics. 

\section{Summary and Outlook}\label{sec:conclusion}

While the dominant mode of gravitational radiation is quadrupolar, higher modes can contribute to the signal when the binary is highly asymmetric or is in highly inclined orbits. Neglecting the effect of higher modes in the parameter estimation of such binaries can cause considerable systematic biases in the inferred source parameters. Fortunately, such binaries are expected to be rare in the detected population, due to (expected) population properties and detector selection effects. However, one might wonder whether neglecting higher modes will affect the inference of \emph{population} properties because this effectively involves the stacking of posteriors from individual events (and hence potentially adding up small biases).

In this work, we study the impact of neglecting higher-order modes in inferring population properties of BBH mergers. We find that, \emph{if} the detected population of BBH mergers does not harbour a significant fraction of asymmetric systems (as predicted by standard population models), the inference of population hyper-parameters will not be biased due to neglecting subdominant modes, assuming an observing scenario with Advanced LIGO-Virgo detectors (O4) containing not more than $\sim 750$ observable events.  It turns out that the accumulation of small biases over many mildly asymmetric systems is not significant enough to bias the inference of hyper-parameters. However, including the effect of higher modes in the parameter estimation will lead to smaller statistical uncertainties on hyper-parameters. Additionally, if the detected population has a significant number of events that have detectable contributions from higher modes, it can bias the maximum mass $m_{\rm{max}}$ of the BH, mass-ratio power-law spectral index $\beta_q$, and the merger rate evolution spectral index $\lambda_z$. This can limit our understanding of various astrophysical implications such as the existence of an upper mass gap in the BH mass spectrum, formation mechanisms of asymmetric compact binary mergers, and evolution of BBH merger rate as a function of redshift. 

We list some caveats of this work:  First, we neglected the effect of BH spins in this study. There are known correlations between the BH spin and mass ratio of the binary, which can reduce the precision in the estimation of hyper-parameters describing the mass ratio distribution of binaries. Second, we assume that the detector data is Gaussian, neglecting the effect of non-Gaussian tails in the data. Third, our conclusions are primarily based on simulation studies using one set of hyper-parameters. Fourth, we consider only one realization of the simulated catalog of BBH mergers and neglect the Poisson fluctuations expected in different realizations. Despite these limitations, we believe that this study provides a useful first understanding of the effect of neglecting higher modes in the population inference. 


The contribution of higher modes is even more pronounced when the binary system has a high and misaligned spin, and when the orbit is non-circular (eccentric). We plan to do a follow-up analysis on the population inference of spinning BBH mergers and see if the inference of population properties improves with the inclusion of subdominant modes in the analysis. An unbiased inference of spin distribution is crucial for understanding the formation channels of compact binary mergers. Another parameter of interest could be eccentricity which could induce a bias in the inference of population properties if not accounted for in waveform models. We leave the study of the effect of neglecting eccentricity in waveform models on the estimation of hyper-parameters for future work.  

\software{\texttt{numpy} \citep{numpy}, \texttt{scipy} \citep{scipy}, \texttt{matplotlib} \citep{matplotlib}, \texttt{pandas} \citep{pandas}, \texttt{astropy} \citep{astropy1, astropy2, astropy3}, \texttt{jupyter} \citep{jupyter}, \texttt{bilby} \citep{bilby}, \texttt{gwpopulation} \citep{gwpopulation}, \texttt{pycbc} \citep{pycbc}, \texttt{corner} \citep{corner}, \texttt{dynesty} \citep{dynesty_sampler}}, \texttt{pesummary} \citep{pesummary}.

\section*{Acknowledgements}
We are grateful to Maya Fishbach and the anonymous referee for their careful review of the manuscript and useful comments. We thank Colm Talbot for the help with running hierarchical inference using \texttt{GWPopulation} \citep{gwpopulation} and estimating selection effects. We are also grateful to Tejaswi Venumadhav and Javier Roulet for illuminating discussions on the project during their visit to ICTS. Our research was supported by the Department of Atomic Energy, Government of India, under project no. RTI4001. PA’s research was, in addition, supported by the Canadian Institute for Advanced Research through the CIFAR Azrieli Global Scholars program. Numerical calculations reported in this paper were performed using the Alice cluster at ICTS-TIFR Bengaluru India and CIT cluster at Caltech, USA. This material is based upon work supported by NSF's LIGO Laboratory which is a major facility fully funded by the National Science Foundation.

\bibliography{references}
\end{document}